\documentclass[sigconf]{acmart}
\AtBeginDocument{%
  }

\copyrightyear{2025} 
\acmYear{2025} 
\setcopyright{cc}
\setcctype{by}
\acmConference[MM '25]{Proceedings of the 33rd ACM International Conference
on Multimedia}{October 27--31, 2025}{Dublin, Ireland}
\acmBooktitle{Proceedings of the 33rd ACM International Conference on
Multimedia (MM '25), October 27--31, 2025, Dublin,
Ireland}
\usepackage{balance} 
\usepackage{enumitem}
\begin{document}

\title{MAGNeT: Multimodal Adaptive Gaussian Networks for Intent Inference in Moving Target Selection across Complex Scenarios}

\author{Xiangxian Li}
\orcid{0000-0001-6638-2361}
\affiliation{%
  \institution{Shandong University}
  \city{Weihai}
  \country{China}
}
\email{xiangxianli@sdu.edu.cn}

\author{Yawen Zheng}
\orcid{0009-0002-3048-4360}
\authornote{Corresponding author.}
\affiliation{%
  \institution{Institute of Software Chinese Academy of Sciences}
  \city{Beijing}
  \country{China}}
\affiliation{%
  \institution{School of Software, Shandong University}
  \city{Jinan}
  \country{China}}
\email{zhengyawen@iscas.ac.cn}

\author{Baiqiao Zhang}
\orcid{0009-0009-7856-5072}
\affiliation{%
  \institution{Shandong University}
  \city{Weihai}
  \country{China}}
\affiliation{%
  \institution{The Hong Kong University of Science and Technology}
  \city{Hong Kong}
  \country{China}}
\email{baiqiao.zhang@connect.ust.hk}

\author{Yijia Ma}
\orcid{0009-0005-6875-2844}
\affiliation{%
  \institution{Shandong University}
  \city{Weihai}
  \country{China}
}
\email{mayijia@mail.sdu.edu.cn}

\author{Xianhui Cao}
\orcid{0009-0007-8908-419X}
\affiliation{%
  \institution{AiLF Instruments}
  \city{Weihai}
  \country{China}
}
\email{hans@ailf.com.cn}

\author{Juan Liu}
\orcid{0000-0003-1069-2875}
\affiliation{%
  \institution{Shandong University}
  \city{Weihai}
  \country{China}
}
\affiliation{%
  \institution{Shandong Key Laboratory of Intelligent Electronic Packaging Testing and Application}
  \city{Weihai}
  \country{China}
}
\email{zzzliujuan@sdu.edu.cn}

\author{Yulong Bian}
\orcid{0000-0003-0999-0656}
\affiliation{%
  \institution{Shandong University}
  \city{Weihai}
  \country{China}
}
\affiliation{%
  \institution{Shandong Key Laboratory of Intelligent Electronic Packaging Testing and Application}
  \city{Weihai}
  \country{China}
}
\email{bianyulong@sdu.edu.cn}

\author{Jin Huang}
\orcid{0000-0002-2833-8041}
\affiliation{%
  \institution{Institute of Software Chinese academy of sciences}
  \city{Beijing}
  \country{China}
}
\email{huangjin@iscas.ac.cn}

\author{Chenglei Yang}
\orcid{0000-0002-9353-8218}
\affiliation{%
  \institution{School of Software, Shandong University}
  \city{Jinan}
  \country{China}
}
\email{chl_yang@sdu.edu.cn}
\renewcommand{\shortauthors}{Xiangxian Li et al.}

\begin{abstract}
  Moving target selection in multimedia interactive systems faces unprecedented challenges as users increasingly interact across diverse, dynamic contexts—from live streaming in moving vehicles to VR gaming in varying environments. Existing approaches rely on probabilistic models that relate endpoint distribution to target properties (size, speed). However, these methods require substantial training data for each new context and lack transferability across scenarios, limiting their practical deployment in diverse multimedia environments where rich multimodal contextual information is readily available. This paper introduces MAGNeT (Multimodal Adaptive Gaussian Networks), which addresses these problems by combining classical statistical modeling with context-aware multimodal method. MAGNeT dynamically fuses pre-fitted Ternary-Gaussian models from various scenarios based on real-time contextual cues, enabling effective adaptation with minimal training data while preserving model interpretability. We take experiments on self-constructed 2D and 3D moving target selection datasets under in-vehicle vibration conditions. Extensive experiments demonstrate that MAGNeT achieves lower error rates with few-shot samples, by applying context-aware fusion of Gaussian experts from multi-factor conditions.
\end{abstract}

\begin{CCSXML}
<ccs2012>
   <concept>
       <concept_id>10010147.10010257</concept_id>
       <concept_desc>Computing methodologies~Machine learning</concept_desc>
       <concept_significance>500</concept_significance>
       </concept>
 </ccs2012>
\end{CCSXML}

\ccsdesc[500]{Computing methodologies~Machine learning}


% \keywords{Do, Not, Us, This, Code, Put, the, Correct, Terms, for,
\keywords{Moving Target Selection, Gaussian Mixture Model, Multi-modal Fusion, Few-shot Learning}


\maketitle

\section{Introduction}

Moving target selection has become increasingly prevalent in multimedia interactive content, such as interactive live streaming and gaming \cite{ilich2009moving}, and is studied as a fundamental task for generalized interfaces (e.g., 2D touchscreens \cite{Huang20192D,Huang2020Crossing} and 3D virtual reality spaces \cite{Zheng3D,Huang2022motion-in-depth}). This task poses significant challenges to users’ perception-action systems \cite{shadmehr2010error,Huang20181D}, especially when performed under complex scenarios involving human factors and environmental perturbations \citep{Dodd2014Touch,mansfield2005effect,kim2013biodynamic}. For instance, Figure \ref{fig:intro} illustrates a scenario where a user watches a live stream on a tablet while traveling in a car. The vehicle's movement makes it challenging to accurately tap on a real-time comment when it scrolls. This leads to the user selecting an unintended comment rather than the intended one. Such errors significantly degrade both user experience and interaction efficiency. Therefore, improving the efficacy of moving target selection amid a combination of human, environmental, and device-related factors.

Previous work has primarily focused on inferring user intent by modeling uncertainties arising from device characteristics, target properties, and human factors. Huang et al. proposed the Ternary-Gaussian model \citep{Huang20181D}, which relates endpoint distributions to a target’s spatial and motion properties. This model provides a statistical criterion for the Bayesian framework, enabling the inference of user intent in moving target selection tasks \citep{Huang20192D,ZhuBayesCommand}.
The model has been expanded to be applied in other scenarios by either incorporating additional impactful factors into the Ternary-Gaussian framework or embedding the Ternary-Gaussian as a kernel within another mechanism \citep{Huang20192D,Huang2022motion-in-depth,Zheng3D,Huang2020Crossing,Zhang2020shape,zhang2023shape,zheng2021AScenario, huang2019modeling, schneider2023supporting} .
However, probabilistic models require substantial data for model fitting, and existing models—often tailored to single contexts—lack transferability; each new scenario demands collection of new data and model parameters.

To address these limitations, this paper formulates the problem of intent inference for moving target selection under complex scenarios, and proposes a Multimodal Adaptive Gaussian Network for Target selection (MAGNeT)\footnote{The project page is \url{https://yibuxulong.github.io/MAGNeT_project/}.}. The proposed MAGNeT effectively senses multimodal environmental data and enables few-shot model adaptation by dynamically fusing pre-fitted Ternary-Gaussian models from various previous scenarios. MAGNeT incorporates user profiling, sensor data, and contextual information of selection tasks, uses a Gaussian mixture model to combine prior models, and applies a self-adaptive learning mechanism to adjust expert model parameters for each context. This significantly reduces prediction error even with limited training samples.

Extensive experiments using self-collected datasets of moving target selections under in-vehicle vibration conditions shows that, MAGNeT achieves lower error rate with only a few-shot samples per user in each condition. Further ablation experiments confirm the effectiveness of combining prior experts. In case studies, MAGNeT adaptively adjusts the weights for environmental factors, and demonstrates strong adaptability. Our key contributions include:

\begin{figure}[t]
    \centering
    \includegraphics[width=0.9\linewidth]{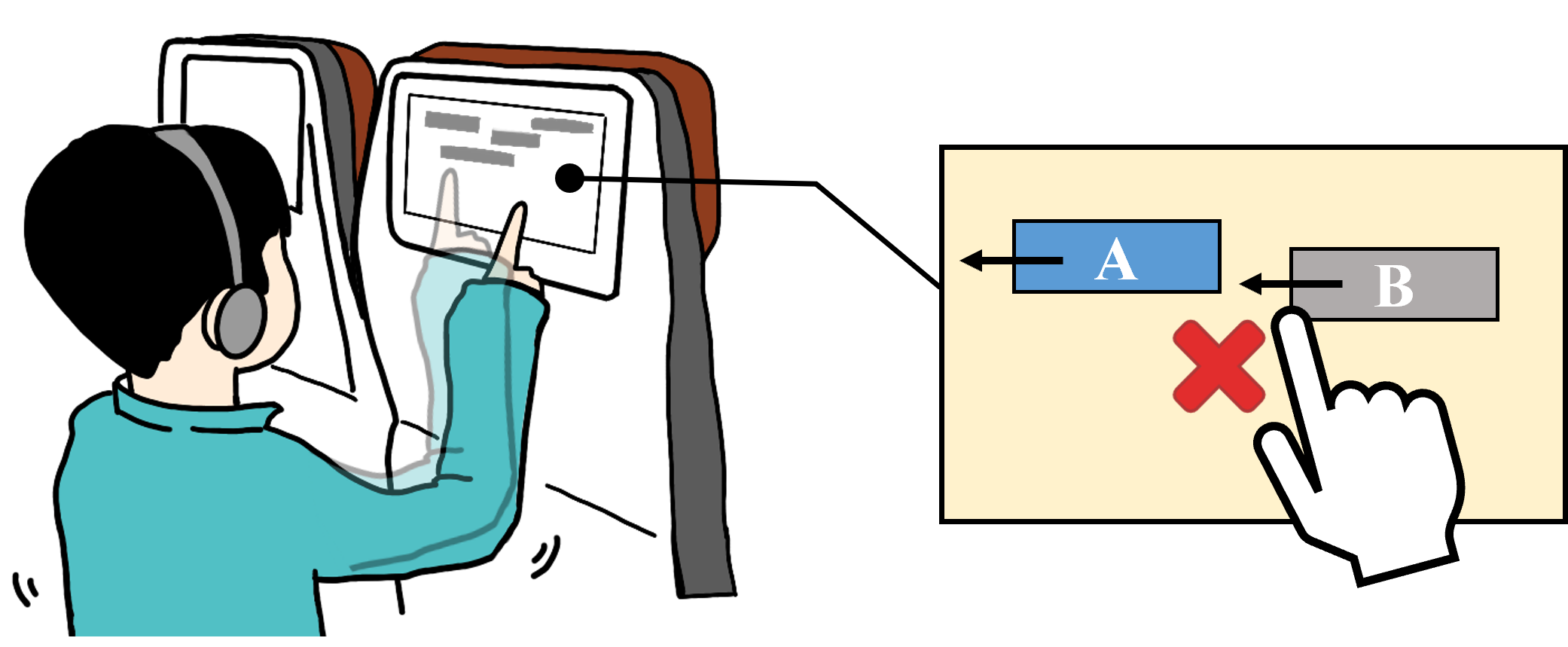}
    \caption{A user is performing touch interaction in a bumpy vehicle cabin. During this process, the car's bumping causes their arm to sway, leading them to accidentally select target B instead of the intended target A.}
    \label{fig:intro}
\end{figure}

\begin{itemize}[leftmargin=10pt]
\item We formulate the problem of intent inference for moving target selection in complex scenarios, and propose MAGNeT, a framework leveraging multimodal context-aware weighting and multi-expert Gaussian modeling.
\item We introduce a self-adaptive update mechanism based on multimodal context information and multi-expert models, adapting prior knowledge while maintaining interpretability.
\item We construct datasets for moving target selection in complex contexts and demonstrate that our approach can dynamically adjust weights to reduce errors even in few-shot settings.
\end{itemize}

\section{Related Works}
\subsection{Modeling Moving Target Selection Behavior}
To enhance user performance in moving target selection, researchers have, on one hand, conducted in-depth studies on explicit assistance methods with visual cues \citep{Vanacken2007BubbleCursor,lu2020investigating,Hasan2011Comet}, and on the other hand, focused on implicit approaches—such as inferring user intent through probabilistic computational models—which leverage interaction behavior uncertainty\citep{ZhuBayesCommand}. Thus, characterizing the uncertainty of moving target selection has become a key concern in the human-computer interaction (HCI) field.
One of the most widely used models is the Ternary-Gaussian model \citep{Huang20181D}. It assumes that the selection endpoint follows a Gaussian distribution composed of three Gaussian components associated with target size, speed, and absolute pointing accuracy.
The Ternary-Gaussian model provides statistical criteria for understanding user interaction intentions by using Bayes' theorem to determine touch selection targets \citep{Huang20192D}.
The Ternary-Gaussian model has been extended from 1D to 2D and 3D space \citep{Huang20192D,Zheng3D} and in various contexts including modeling path steering and pursuit, crossing-based moving target selection \citep{Huang2020Crossing}, endpoint distribution of arbitrary shapes \citep{Zhang2020shape,zhang2023shape}, touching in different scenarios \citep{zheng2021AScenario}, and spatiotemporal selection \citep{huang2019modeling,schneider2023supporting}.
\subsection{Gaussian Mixture Model for Modeling Uncertainty}
Existing models analyze individual factors like target size, speed, and depth influencing moving target selection. However, these factors combine non-linearly, making it hard for single models to capture their coupled effects. Gaussian Mixture Models (GMMs), widely used for interaction uncertainty modeling \cite{mclachlan2000finite,fang2023uncertainty}, approximate complex probability distributions. For example, GMMs improve robot behavior prediction by integrating environmental disturbances and user intent. Yet, current GMMs rely on static weights, struggling with real-time changes in dynamic interactions. This points to a need for context-aware dynamic weight calibration. This would involve real-time sensing of environmental fluctuations, user operations, and target motion (e.g., speed and size) to adaptively adjust expert model weights, overcoming GMM limitations in dynamic coupling. In dynamic systems, Mixture of Experts (MoE) models use gating networks to adjust sub-model weights, providing a framework for multi-factor coupling \cite{li2023adaptive,yi2024variational}. However, traditional MoE models depend on offline data, limiting real-time adaptability to contextual changes \cite{cao2023multi,gan2025mixture}. Bayesian online adaptive methods can update weights but face high computational costs for high-dimensional features and lack explicit modeling of factor coupling \cite{gershman2012tutorial}.

\section{Problem Formulation}
In the scenario of moving target selection based on the Ternary-Gaussian model, taking 2D interface as an example, each target $t_i \in T$ is defined by spatial attributes (screen coordinates ($x_i$,$y_i$), target size $w_i \in W$) and motion attributes (velocity $v_i \in V$, moving direction $\alpha_i$). Using the coordinates $(x_s, y_s)$ of a user’s interaction endpoint $s$ and its relative state to targets as input, the method leverages pre-fitted parameters $\theta_{init}$ derived from large-scale empirical datasets to initialize the model, requiring at least 9 predefined pairs of Gaussian parameters ($\mu$ and $\Sigma$) corresponding to distinct $W \times V$ conditions and a minimum of 100 endpoint samples per condition for robust bivariate distribution estimation. For each target ${t_i}$, the model predicts its endpoint distribution parameters  ($\mu_i$ and $\Sigma_i$) based on its attributes, computes the likelihood $p_{i}$ for an observed endpoint $s$, and applies Bayesian inference to determine the posterior probability, ultimately selecting the target with the highest posterior probability as the user’s intended choice. 

However, existing methods not only require large-scale datasets for initial model construction but also employs fixed parameters, thereby limiting adaptability to novel interaction contexts. Considering that multiple Ternary-Gaussian expert models (${\theta_1,...,\theta_k}$) derived from diverse acquisition devices or user populations—exhibit inherent variability in parameter ranges and inter-parameter relationships, their generalization abilities in new environments are uncertain. Different from the previous learning paradigm, we propose MAGNeT as an adaptive parameter adjustment strategy that tailors model parameters to current settings. This is complemented by context-aware feature extraction, which captures intricate situational characteristics to facilitate fusion of Ternary-Gaussian expert models within a Gaussian mixture framework. Our approach enhances selection performance under small sample conditions while retaining the interpretability advantages of the original Ternary-Gaussian model framework.

\begin{figure*}[t]
  \centering
  \includegraphics[width=0.8\linewidth]{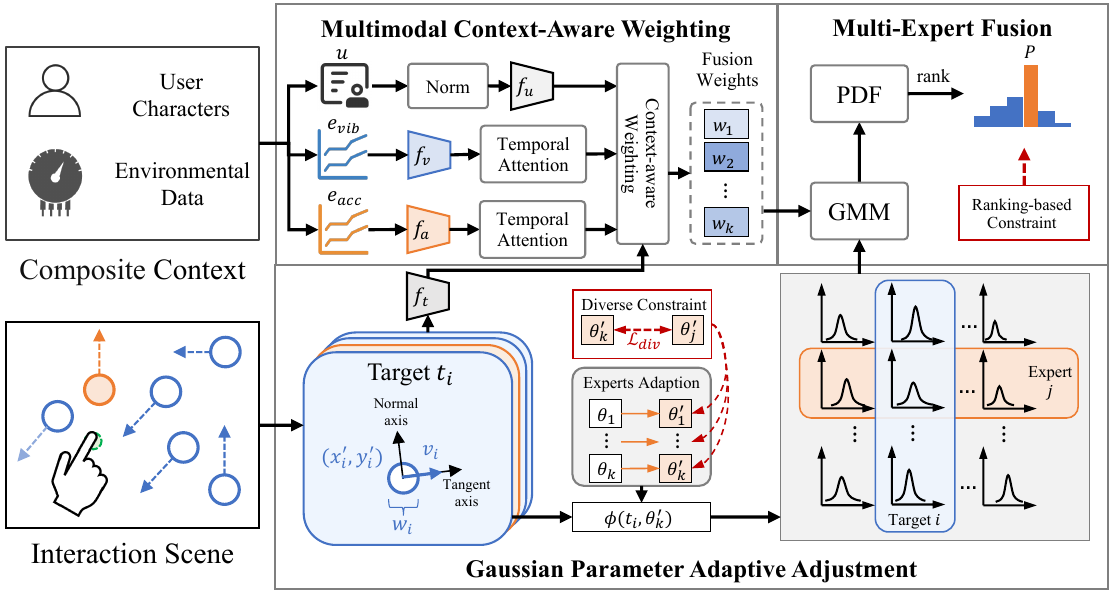}
  \caption{Illustrative diagram of MAGNeT, which features a Multimodal Context-Aware Weighting module that deciphers how prior models collaborate in a given scene. And its Gaussian Parameter Adaptive Adjustment module refines prior parameters specifically for the current context. The resulting ensemble of Gaussian distributions predicts outcomes via Gaussian mixture.}
  \label{fig:framework}
\end{figure*}

\section{Methodology}
\label{sec:methodology}

To address the issues of intent inference under complex interaction scenario, this paper proposes a \textbf{ Multimodal Adaptive Gaussian Network (MAGNeT)}. As illustrated in Figure \ref{fig:framework}, MAGNeT consists of three components: (1) Multimodal Context-Aware Weighting module captures and fuses heterogeneous contextual information, (2) Gaussian Parameter Adaptive Adjustment module dynamically adapts Gaussian parameters based on target-specific contexts, and (3) Multi-expert Fusion module generates probabilistic touch predictions through Gaussian mixture modeling.

\subsection{Multimodal Context-Aware Weighting}
\label{subsec:context_weighting}

The multimodal context-aware weighting module serves as the foundation of our framework, employing specialized encoders to extract features from different modalities and generating adaptive fusion weights for expert selection.

\subsubsection{User Feature Encoder}
User characteristics including gesture type, age, and gender are first normalized and then processed through a dedicated user encoder $f_u$:
\begin{equation}
\mathbf{h}_u = f_u(\text{Norm}(\mathbf{u}))
\end{equation}
where $\mathbf{u}$ represents the raw user features and $f_u$ is the user encoder that transforms normalized user characteristics into a latent representation $\mathbf{h}_u \in \mathbb{R}^{d_{u}}$, where $d_{u}$ is the dimension of user characters.

\subsubsection{Environmental Feature Encoders}
Environmental data from vibration and acceleration sensors are processed through specialized encoders with temporal attention mechanisms:

For vibration signals $\mathbf{e}_{vib} \in \mathbb{R}^{T \times d_{v}}$:
\begin{equation}
\mathbf{h}_v = f_v(\text{TemporalAttention}(\mathbf{e}_{vib}))
\end{equation}

For acceleration signals $\mathbf{e}_{acc} \in \mathbb{R}^{T \times d_{a}}$:
\begin{equation}
\mathbf{h}_a = f_a(\text{TemporalAttention}(\mathbf{e}_{acc}))
\end{equation}
where $f_v$, $f_a$, $d_v$ and $d_a$are the vibration encoder, acceleration encoder, vibration dimension, and acceleration dimension, respectively. The temporal attention mechanism allows the model to focus on the most relevant time steps for each environmental modality before feature encoding.

\subsubsection{Target Feature Encoder}
The target-specific information from the interaction scene is processed through a target encoder $f_t$:
\begin{equation}
\mathbf{h}_t = f_t(\mathbf{t}_i)
\end{equation}
where $\mathbf{t}_i$ contains the target information including spatial coordinates and geometric properties, and $f_t$ is the target feature encoder.

\subsubsection{Context-Aware Weighting}
All encoded representations are integrated through a context-aware weighting (CAW) model to generate expert fusion weights:
\begin{align}
\mathbf{h}_{con} &= \text{Concat}(\mathbf{h}_u, \mathbf{h}_v, \mathbf{h}_a, \mathbf{h}_t) \\
\mathbf{w} &= [w_1, w_2, \ldots, w_k] = \text{CAW}(\mathbf{h}_{con})
\end{align}
where $\mathbf{w} \in \mathbb{R}^k$ represents the fusion weights for $k$ experts, and $\sum_{i=1}^k w_i = 1$. The CAW architecture consists of three linear layers with LeakyReLU activations, where batch normalization and dropout are applied after the first layer.

\subsection{Gaussian Parameter Adaptive Adjustment}
\label{subsec:adaptive_adjustment}

This module performs target-specific Gaussian parameter adaptation, enabling the model to dynamically adjust to different touch targets and interaction contexts.

\subsubsection{Target-Specific Coordinate System}
For each target $t_i$, we establish a coordinate system with normal and tangent axes based on the target's geometric properties:
\begin{equation}
\mathbf{t}_i = \{(x'_i, y'_i), \mathbf{v}_i, \mathbf{w}_i\}
\end{equation}
where $(x'_i, y'_i)$ are the transformed coordinates, $\mathbf{v}_i$ is the tangent axis, and $\mathbf{w}_i$ is the normal axis.

\subsubsection{Expert Parameter Adaptation}
We maintain $k$ expert parameter sets $\{\boldsymbol{\theta}_1, \boldsymbol{\theta}_2, \ldots, \boldsymbol{\theta}_k\}$. Each expert's parameters are adapted based on the target-specific context and encoded features:
\begin{equation}
\boldsymbol{\theta}'_k = \text{ExpertAdaptation}(\boldsymbol{\theta}_k, \mathbf{h}_t, \mathbf{h}_{con})
\end{equation}

The expert adaptation process leverages both the target-specific features $\mathbf{h}_t$ from encoder $f_t$ and the global context $\mathbf{h}_{con}$.

\subsubsection{Diversity Constraint}
To ensure expert diversity and prevent parameter collapse, we introduce a diversity constraint $\mathcal{L}_{div}$:
\begin{equation}
\mathcal{L}_{div} = \frac{1}{k(k-1)} \sum_{i=1}^k \sum_{j \neq i} \text{sim}(\boldsymbol{\theta}'_i, \boldsymbol{\theta}'_j)
\end{equation}
where $\text{sim}(\cdot, \cdot)$ measures the similarity between expert parameters, encouraging diverse specialization across experts.

\subsubsection{Gaussian Distribution Generation}
For each expert $k$ and target $t_i$, we generate the corresponding Gaussian distribution:
\begin{equation}
\mathcal{G}_{k,i} = \phi(t_i, \boldsymbol{\theta}'_k)
\end{equation}
where $\phi(\cdot, \cdot)$ is the Gaussian parameterization that computes the mean and covariance matrix based on the adapted parameters.

\subsection{Multi-expert Fusion}
\label{subsec:multi_expert_fusion}

The final module combines predictions from multiple experts through Gaussian mixture to generate the final probability distribution.

\subsubsection{Gaussian Mixture Model Construction}
The Gaussian mixture model combines expert predictions weighted by the context-aware fusion weights derived from the encoded features:
\begin{equation}
p(\mathbf{y}|t_i, \mathbf{h}_{con}) = \sum_{k=1}^K w_k \mathcal{G}_{k,i}(\mathbf{y})
\end{equation}
where $\mathbf{y} \in \mathbb{R}^2$ represents the 2D touch coordinates, and the weights $w_k$ are computed from the concatenated encoded features $\mathbf{h}_{con}$.

\subsubsection{Probability Density Function}
The GMM outputs a probability density function (PDF) that captures the uncertainty of touches:
\begin{equation}
\text{PDF}(t_i) = p(\mathbf{y}|t_i, \mathbf{h}_{con})
\end{equation}

\subsubsection{Ranking-based Prediction}
The final prediction is obtained by ranking targets based on their probability values:
\begin{equation}
\text{rank}(t_i) = \arg\max_{\mathbf{y}} \text{PDF}(t_i)
\end{equation}

\subsection{Loss Function and Training Strategy}
\label{subsec:loss_training}

Our training objective combines two constraints:
\begin{equation}
\mathcal{L}_{total} = \mathcal{L}_{rank} + \lambda_{div} \mathcal{L}_{div}
\end{equation}

\subsubsection{Ranking-based Constraint}
The ranking loss ensures that the ground truth target has higher probability than negative samples:
\begin{equation}
\mathcal{L}_{rank} = \max(0, \text{margin} + \log p(\mathbf{y}_{neg}|t_{neg}) - \log p(\mathbf{y}_{pos}|t_{pos}))
\end{equation}
where $t_{pos}$ is the ground truth target and $t_{neg}$ represents negative targets. The complete framework effectively integrates multimodal contextual information through specialized encoders ($f_u$, $f_v$, $f_a$, $f_t$) and adaptive parameter adjustment.

\section{Data Collection}

\subsection{Participants}
 10 participants (3 females) were recruited in this study. The average age of the participants was 23.4 years old $(\pm$ 2.84), and all of them were right-handed. Every participant had experience in using touch-based devices such as smartphones and tablets. None of the participants reported having any perceptual or motor impairments.
\subsection{Apparatus}
\label{sec:Apparatus}
This study collected environmental vibration and user interaction data within a real vehicle on a 2-km closed campus loop to minimize external interference and ensure safety. The route featured 6 right-angle turns and 8 speed bumps. The weighted root mean square acceleration (RMSA) is used to characterize the intensity of the vibration, in accordance with the ISO 2631–1 standard (International Organization for Standardization, 1997 \citep {an1997mechanical}). Due to non-regular, aperiodic vibrations, RMSA was calculated over 3-second windows preceding each target selection.
\begin{figure}[t]
    \centering
    \includegraphics[width=0.95\linewidth]{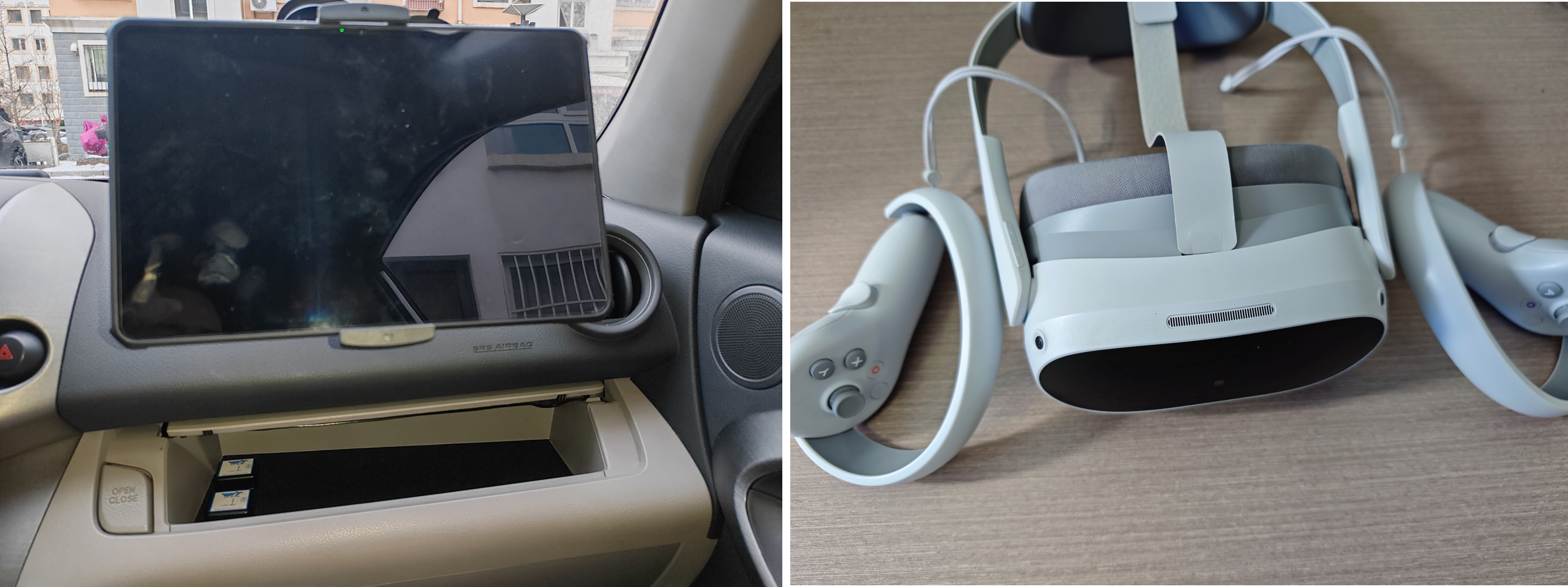}
    \caption{Apparatus used in the experiment.}    
    \label{fig:device}
\end{figure}

During the experiment, the experimenter drove a car (Toyota RAV4) along the specified experimental route. Environmental vibration was measured using two inertial sensors (Wit BWT901BLECL5.0, Wit WTVB01-BT50) mounted on the passenger-side platform. They were used to collect acceleration data and vibration data respectively. The acceleration data included the acceleration of the x, y, and z axes, and the vibration data included the vibration velocity, vibration angle, vibration displacement, and vibration frequency of the x, y, and z axes. As shown in Figure \ref{fig:device}, A Huawei MatePad Pro tablet (2560×1600, 12.6", 240 PPI) displayed 2D target selection tasks and recorded touch points, mounted near-vertially on a bracket. A Pico 4 headset (4320×2160, 2.56" per eye, 1200 PPI, 105° FOV) presented 3D selection tasks and captured spatial pointing data.

\subsection{Design}
\subsubsection{2D moving target selection}
The experiment adopted a within-subjects design, involving a cross-combination of 4 target size levels, 4 target speed levels, and 2 interaction gestures:  
\begin{itemize}[leftmargin=10pt]
\item Target size ($W$): 65, 95, 125, and 155 px
\item Target speed ($V$): 300, 550, 800, and 1050 px/s
\item Interaction gesture ($P$): tablet fixed, tablet handheld
\end{itemize}
Each participant completed 12 repeated trials under each experimental condition, totaling $W$(4) × $V$(4) × $P$(2) × 10 participants × 12 repetitions = 3840 trials. 
\begin{figure}[t]
    \centering
    \includegraphics[width=1\linewidth]{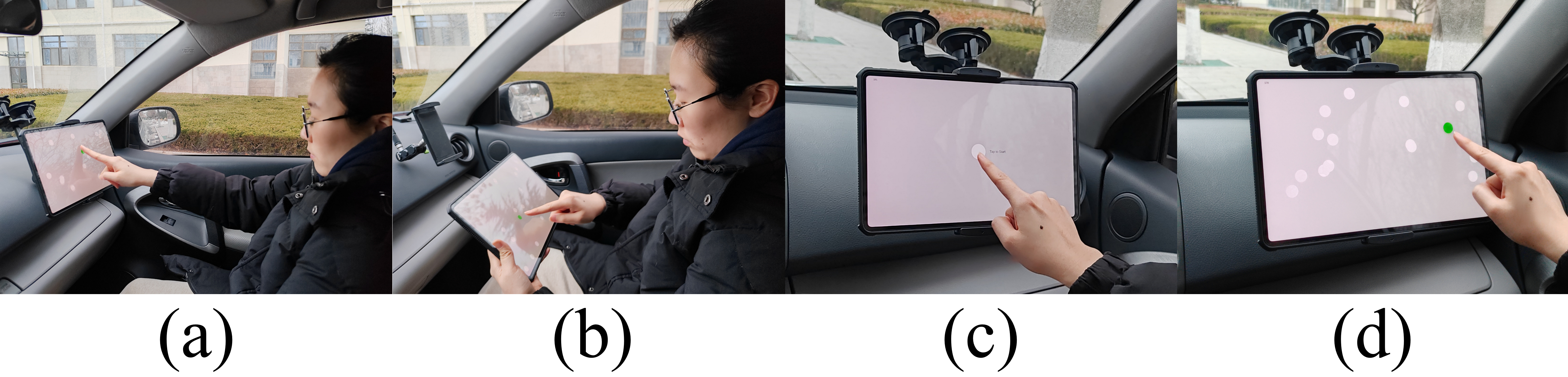}
    \caption{Two interactive gestures and task interface of 2D moving target selection.}
    \label{fig:2Dtask-new}
\end{figure}
During the experiment, participants adopted two interaction gestures: the tablet fixed in the car (Figure \ref{fig:2Dtask-new}(a)) and the tablet handheld by the participant (Figure \ref{fig:2Dtask-new}(b)). The order of interaction gestures and the presentation sequence of target conditions were counterbalanced across participants, with rest periods inserted between different conditions. Additionally, the test order within each condition was randomized. Each participant took approximately 25 minutes to complete the entire test.

\subsubsection{3D moving target selection}
Within-subjects design was employed in the experiment, involving crossed combinations of 4 target size levels and 4 target speed levels:  
\begin{itemize}[leftmargin=10pt]
\item Target size ($W$): 0.04, 0.08, 0.12, and 0.16 m  
\item Target speed ($V$): 0.22, 0.34, 0.45, and 0.56 m/s  
\end{itemize}
Each participant completed 6 repeated trials under each experimental condition, totaling $W$(4) × $V$(4) × 10 participants × 6 repetitions = 960 trials. The presentation sequence of target conditions was counterbalanced across participants, with rest intervals inserted between different conditions. Additionally, the test order within each condition was randomized. One participant took approximately 15 minutes to complete the entire experiment.
\subsection{Procedure}
\subsubsection{2D moving target selection}
\begin{figure}[t]
    \centering
    \includegraphics[width=\linewidth]{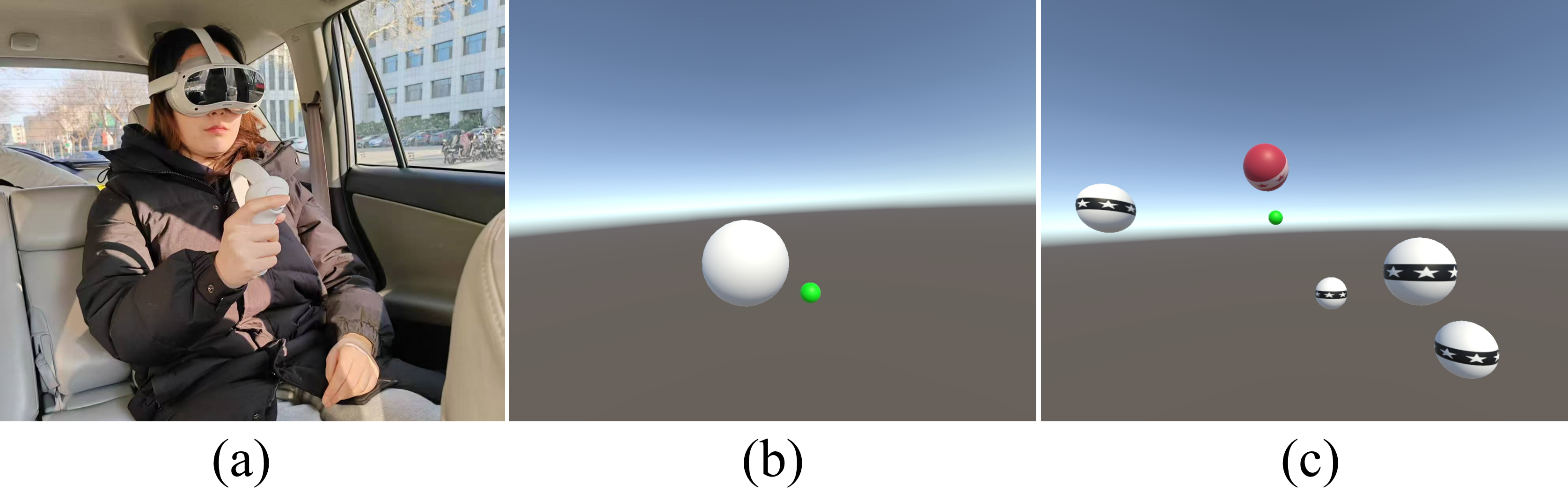}
    \caption{Interactive gesture and task interface of 3D scenario.}
    \label{fig:3Dtask}
\end{figure}

Participants sat in the passenger seat, adjusted the seat to a comfortable position, and fastened their seatbelt. Once ready, the experimenter started the vehicle and drove along the predetermined route. In the 2D moving target selection task, participants initiated the test by tapping the "Start" button at the center of the screen, as shown in Figure \ref{fig:2Dtask-new} (c). At the beginning of each trial, 15 circular targets appeared at random positions on the screen and moved in random directions at a fixed speed. If a target reached the screen edge, it rebounded according to the law of specular reflection while maintaining its speed. Among these 15 circular targets, 14 were white, and only one was specially marked green, as shown in Figure \ref{fig:2Dtask-new} (d). Participants were instructed to tap the green target on the screen with their finger as quickly and accurately as possible. Each participant had only one attempt per task, and the system recorded the specific coordinates of their touch point on the screen regardless of whether the tap successfully hit the target's interior. The screen would then automatically clear, and the "Start" button would reappear at the center. Participants needed to tap this button again to initiate the next trial. When the vehicle completed one lap along the predetermined route and returned to the starting point, participants took a full rest before switching interaction gestures for further testing.
\subsubsection{3D moving target selection}
Participants made selection operations by squeezing the trigger of a handle, with the handle real-time mapped to a cursor in the 3D scene. The 3D cursor, a green sphere with a diameter of 1 cm, was dynamically aligned with the handle's physical position in the real world, allowing participants to determine the handle's location by observing the green sphere. Participants initiated the test by selecting a white sphere at the center of their field of view (Figure \ref{fig:3Dtask} (b)).

At the start of each trial, 5 spherical targets appeared at random positions in the 3D space in front of the participant, moving in random directions at a fixed speed. The targets' movement range had a depth variation interval of 0.25 m to 0.6 m. Among the 5 spherical targets, only one was marked red (Figure \ref{fig:3Dtask} (c)). Participants were instructed to use the handle to select the red target as quickly and accurately as possible.  

Each participant had only one attempt per task, and the system recorded the specific spatial coordinates of the 3D cursor regardless of whether the selection successfully hit the target's interior. Subsequently, all targets were automatically cleared, and a white sphere reappeared at the center of the field of view. Participants could initiate the next trial by selecting this sphere again.

\begin{table}[t]
\centering
\caption{Statistics of datasets, \# means the number of item.}
    \begin{tabular}{ccccc}
    \toprule
       \textbf{Dataset} & \textbf{\# Samples} & \textbf{\# Targets} & \textbf{\# Test}&  \textbf{\# Validation} \\
       \midrule
        MTS-2D &  3,840&  15& 1,536 &  384 \\
        MTS-3D &  960&  5&  384&  96 \\
        \bottomrule
    \end{tabular}
    \label{tab:dataset}
\end{table}

\begin{table*}[t]
\centering
\caption{Model performances on MTS-2D dataset. We report means (standard deviations in parentheses) across different seeds.}
\label{tab:compare_2d}
\begin{tabular}{lcccccccc}
\toprule
\textbf{Model} & \textbf{$E_{clust}(G1)$} & \textbf{$E_{clust}(G2)$} & \textbf{$E_{mean}(G1)$} & \textbf{$E_{mean}(G2)$} & \textbf{$E@1$} & \textbf{$E@2$} \\
\midrule
Border-based   & 0.7844(0.0068) & 0.8318(0.0080) & 0.7694(0.0109) & 0.8307(0.0050) & 0.8003(0.0042) & - \\
Distance-based     & 0.2223(0.0089) & 0.2285(0.0088) & 0.2021(0.0098) & 0.2463(0.0046) & 0.2243(0.0039) & 0.0657(0.0034) \\
Expert(s-f)  & 0.1367(0.0035) & 0.1681(0.0153) & 0.1239(0.0029) & 0.1705(0.0140) & 0.1473(0.0072) & 0.0400(0.0021) \\
Expert(s-h) & 0.1189(0.0053) & 0.1637(0.0152) & 0.1109(0.0041) & 0.1569(0.0121) & 0.1341(0.0077) & 0.0359(0.0018) \\
Expert(w-h) & 0.1156(0.0073) & 0.1558(0.0158) & 0.1061(0.0044) & 0.1521(0.0132) & 0.1292(0.0087) & 0.0349(0.0015) \\
\hline
MAGNeT (1-Shot)   & 0.1158(0.0058) & 0.1584(0.0152) & 0.1069(0.0044) & 0.1533(0.0130) & 0.1303(0.0085) & 0.0357(0.0031) \\
MAGNeT (3-Shot)   & 0.1110(0.0027) & 0.1554(0.0132) & 0.1025(0.0018) & 0.1494(0.0106) & 0.1261(0.0049) & \textbf{0.0341}(0.0019) \\
MAGNeT (5-Shot)   & 0.1122(0.0055) & {0.1535}(0.0141) & 0.1012(0.0028) & 0.1509(0.0130) & 0.1262(0.0074) & 0.0353(0.0016) \\
MAGNeT (10-Shot)  & \textbf{0.1093}(0.0047) & \textbf{0.1524}(0.0128) & \textbf{0.1002}(0.0028) & \textbf{0.1474}(0.0114) & \textbf{0.1239}(0.0064) & 0.0351(0.0030) \\
\bottomrule
\end{tabular}
\end{table*}

\section{Experiments}
\subsection{Experimental Settings}
\subsubsection{Dataset} The experiments are conducted on two datasets. Statistics of datasets are shown in Table \ref{tab:dataset}.

\textbf{Moving Target Selection 2D (MTS-2D)} is a dataset containing 3,840 samples of 2D moving target selection touch data. Of these, 1,536 samples are used for model testing and 384 for validation. 

\textbf{Moving Target Selection 3D (MTS-3D)} is a dataset containing 960 samples of 3D moving target selection data. Of these, 384 samples are used for model testing and 96 for validation.

The dataset partitioning is performed via uniform sampling across target conditions ($W \times V$), respectively. And the train set of each dataset is sampled from the remaining data according to the specific experimental settings described in Section \ref{sec:Comparative Results}. Both datasets include user characteristics (age, gender, gesture) and environmental vibration characteristics introduced in Section \ref{sec:Apparatus}.

\subsubsection{Expert Models}
For the MTS-2D dataset, we have three expert models corresponding to sit-fixed, sit-handheld, and walk-handheld scenarios. We use \textbf{Expert (s-h)}, \textbf{Expert (s-f)}, and \textbf{Expert (w-h)} to denote the models derived from these scenarios, respectively:  
\begin{itemize}[leftmargin=10pt]
    \item Expert (s-f): A 2D Ternary-Gaussian model with parameters fitted from experimental data of 25 participants. Participants remained seated and extended their arms to interact with a fixed tablet in the cabin for moving target selection.  
    \item Expert (s-h): A 2D Ternary-Gaussian model with parameters fitted from experimental data of 21 participants. Participants sat while holding the tablet in one hand and using the other hand to tap the screen for moving target selection tasks.  
    \item Expert (w-h): A 2D Ternary-Gaussian model with parameters fitted from experimental data of 21 participants. Participants walked while holding the tablet in one hand and using the other hand to tap the screen for moving target selection.  
\end{itemize}  

For the MTS-3D dataset, a 3D Ternary-Gaussian model (Expert (3D)) whose parameters were derived from \citep{Zheng3D}.

\subsubsection{Parameter Settings}
For model training, we configure the batch size to 32 for the MTS-2D and 16 for the MTS-3D to ensure stable gradient estimation. The maximum number of training epochs is set to 50, with early stopping triggered if the validation loss fails to decrease for 10 consecutive epochs. We employ the AdamW optimizer for model optimization with a learning rate of $5 \times 10^{-4}$ and weight decay of $1 \times 10^{-4}$ to promote generalization. A cosine annealing schedule is applied for learning rate decay.

Regarding the specific architectural parameters of our MAGNeT framework, the feature encoders are configured as follows:

\textbf{User and Target Encoders:} Both the user encoder $f_u$ and target encoder $f_t$ are implemented as linear networks with hidden dimension $d_{hidden} = 64$, producing user representations $\mathbf{h}_u \in \mathbb{R}^{64}$ and target representations $\mathbf{h}_t \in \mathbb{R}^{64}$, respectively.

\textbf{Environmental Encoders:} The acceleration encoder $f_a$ and vibration encoder $f_v$ are implemented as bidirectional GRU networks, each with hidden dimension $d_{hidden} = 64$, generating environmental feature representations $\mathbf{h}_a, \mathbf{h}_v \in \mathbb{R}^{128}$.

\textbf{Temporal Attention:} The temporal attention mechanism employs multi-head attention with feature dimension 128 and 8 attention heads, enabling the model to capture diverse temporal dependencies in environmental signals.

\textbf{Context-Aware Weighting:} The softmax temperature parameter $\tau$ in the context-aware weighting module is set to 2.0 to control the sharpness of expert weight distributions.

\subsection{Metrics}
To analyze the model's ability to model interaction uncertainty in complex scenarios, this paper measures the error rate of the model's predictions. For the paradigm of secondary confirmation in interaction assistance \cite{Mankoff2000Providing,Li2022Select}, we measured the Top-1 and Top-2 error rates, formed as $E@1$ and $E@2$. Additionally, to model the uncertainty caused by vibrations in an in-vehicle environment, we further measured the error rates under different levels of vibration.
\begin{itemize}[leftmargin=10pt]
    \item \textbf{Top-1 Error rates based on clustering ($E_{clust}$)}: K-means clustering was performed on the acceleration and vibration data collected during the experiment. The number of clusters was determined by the silhouette coefficient. Using RMSA as a quantitative indicator, the midpoint of the clusters was calculated based on the cluster centers to serve as the dividing point. Statistical results show that in the 2D dataset, the data can be divided into two clusters. The mean RMSA of the first and second clusters are 0.7030 and 0.4516 separately. Therefore, for the 2D data, group 1 (G1) for $E_{clust}$ is $RMSA < 0.5773$, and G2 is $RMSA \geq 0.5773$. The 3D dataset is also clustered into two groups: the mean RMSA of clusters are 0.3895 and 0.5768 separately. Therefore, for the 3D data, G1 for $E_{clust}$ is $RMSA < 0.4831$, and G2 is $RMSA\geq0.4831$.
    \item \textbf{Top-1 Error rates based on the mean of RMSA ($E_{mean}$)}: Using the mean of RMSA as the benchmark, in the 2D dataset, G1 for $E_{mean}$ is $RMSA < 0.5508$, and G2 is $RMSA \geq 0.5508$. In the 3D dataset, G1 for $E_{mean}$ is $RMSA < 0.4671$, and G2 is $RMSA \geq 0.4671$.
\end{itemize}

\begin{table*}[t]
\centering
\caption{Model performances on MTS-3D dataset. We report means (standard deviations in parentheses) across different seeds.}
\label{tab:compare_3d}
\begin{tabular}{lcccccccc}
\toprule
\textbf{Model} & \textbf{$E_{clust}(G1)$} & \textbf{$E_{clust}(G2)$} & \textbf{$E_{mean}(G1)$} & \textbf{$E_{mean}(G2)$} & \textbf{$E@1$} & \textbf{$E@2$} \\
\midrule
{Border-based} & 0.8417(0.0146) & 0.8519(0.0108) & 0.8357(0.0170) & 0.8548(0.0164) & 0.8458(0.0097) & - \\
{Distance-based} & 0.4235(0.0288) & 0.3757(0.0397) & 0.4389(0.0297) & 0.3721(0.0387) & 0.4047(0.0294) & 0.2214(0.0219) \\
{Expert(3D)} & 0.3121(0.0261) & 0.2184(0.0350) & 0.3265(0.0225) & 0.2273(0.0317) & 0.2759(0.0168) & 0.1179(0.0152) \\
\hline
{MAGNeT (1-Shot)} & 0.0101(0.0071) & 0.0166(0.0093) & 0.0076(0.0073) & 0.0173(0.0109) & 0.0125(0.0074) & 0.0006(0.0013) \\
{MAGNeT (2-Shot)} & \textbf{0.0042}(0.0062) & \textbf{0.0048}(0.0039) & \textbf{0.0027}(0.0054) & \textbf{0.0061}(0.0054) & \textbf{0.0044}(0.0047) & \textbf{0.0000}(0.0000) \\
\bottomrule
\end{tabular}
\end{table*}

\subsection{Comparative Results}
\label{sec:Comparative Results}
To evaluate the effectiveness of MAGNeT, we conducted comparative experiments against several approaches. 1) The Border-based method determines whether a target is successfully selected based on whether the user’s touch point falls within the radius of the target. This method serves as a standard baseline for moving target selection tasks. 2) The Distance-based method identifies the user’s intended target as the one with the smallest Euclidean distance to the touch point. 3) The Expert models refer to expert models, the specific definitions of which have been detailed in earlier sections.  We compare MAGNeT with few-shot training samples with aforementioned methods, and following the previous works \cite{huang2019modeling,zheng2021AScenario}, we refer 1-shot setting as each participant contributes only one selection sample under each combination of target size ($W$) and speed ($V$). And similar to the 2-shot, 3-shot, 5-shot, and 10-shot settings. Based on the performance comparison results presented in Table \ref{tab:compare_2d} and \ref{tab:compare_3d}, we summarize the key findings as follows:
\begin{itemize}[leftmargin=10pt]
    \item \textbf{Border-based methods yield high error rates in both 2D and 3D datasets, indicating that accurately selecting targets is non-trivial, which highlights the critical importance of incorporating intent understanding to assist in target selection.} As shown in Tables \ref{tab:compare_2d} and \ref{tab:compare_3d}, the Border-based method suffers from substantial error across both datasets, with $E_{mean}$ values as high as 0.7694 (2D) and 0.8357 (3D). These high error rates underscore the difficulty of directly selecting moving targets—especially under complex selection conditions.
    \item \textbf{Distance-based methods achieve partial improvement but struggle in complex interaction scenarios.} Compared with Border-based methods, Distance-based approaches significantly reduce selection error (e.g., $E_{mean}(G1)$ drops from 0.7694 to 0.2021 in 2D). However, their performance remains inferior to Experts, especially in scenarios with environmental noise and user variability. Given the complex dynamics of real-world settings, Euclidean distance proves insufficient to capture user intent, particularly when movement-induced noise and ambiguity are present—highlighting the need for more expressive models.
    \item \textbf{Expert models offer further gains but show limited generalization across contexts.} The Expert models, trained on specific conditions such as walking, sitting, or vibration, perform better than baseline methods. However, the performance across Experts varies (e.g., Expert(s-h) vs. Expert(s-f)), revealing sensitivity to the context in which they were trained. In particular, their generalization to new or unseen contexts is constrained, as no adaptive mechanism is available to adjust their weights based on the current interaction scenario.
    \item \textbf{MAGNeT is effective in low-data regimes and generalizes well across motion conditions.} Across both MTS-2D and MTS-3D datasets, MAGNeT exhibits remarkable learning capabilities. Its performance remains consistent across different random seeds (G1 and G2), with low variance in error metrics. Importantly, this consistency holds under both sparse data (e.g., 1 or 2 samples per condition) and varying interaction dynamics, confirming that its self-calibrating mixture of experts and context encoding mechanisms generalize well across interaction complexities.
    \item \textbf{MAGNeT achieves significant increments on the MTS-3D dataset, validating the effectiveness of adaption.} MAGNeT demonstrates superior performance across metrics in the 3D setting. For instance, the 2-Shot variant achieves an exceptionally low $E_{mean}(G1)$ of 0.0027, outperforming all baselines by a wide margin. Given the challenge of selecting 1 target from 5 under motion, such performance illustrates MAGNeT’s strength in combining contexts with real-time expert adaptation, enabling robust intent prediction even in three-dimensional environments.
    
\end{itemize}

\begin{table}[t]
\centering
\caption{Ablation results. }%The s-f, s-h, w-h means the experts modeling from the sit-fixed, sit-handheld, walk-handheld conditions, respectively.}
\label{tab:ablation_expert}
\begin{tabular}{cccccc}
\toprule
\textbf{Metric} & MAGNeT & w/o s-f & w/o s-h & w/o w-h & w/o all \\
\midrule
{\textbf{$E_{clust}(G1)$}}& 0.1093 & 0.1108 & 0.1101 & 0.1093& 0.1122 \\
{\textbf{$E_{clust}(G2)$}} & 0.1524 & 0.1531 & 0.1513 & 0.1535 & 0.1573 \\
{\textbf{$E_{mean}(G1)$}}    & 0.1002 & 0.1007 & 0.1007 & 0.1002 & 0.1038 \\
{\textbf{$E_{mean}(G2)$}}    & 0.1474 & 0.1494 & 0.1471 & 0.1481 & 0.1509 \\
{\textbf{$E@1$}}          & 0.1239 & 0.1252 & 0.1241 & 0.1243 & 0.1275 \\
{\textbf{$E@2$}}        & 0.0351 & 0.0334 & 0.0351 & 0.0348 & 0.0363 \\
\bottomrule
\end{tabular}
\end{table}

\subsection{Ablation of Experts Selection}
To investigate the impact of expert models collected under diverse scenarios when utilized as initialization values on model performance, we conducted ablation experiments on a 2D dataset with a 10-shot training set. As shown in Table \ref{tab:ablation_expert}, key observations from the results are summarized as follows:
\begin{itemize}[leftmargin=10pt]
\item \textbf{Limited impact of single-expert absence, but significant degradation when all experts are missing.}
The absence of any individual expert model exhibits a marginal effect on overall performance, demonstrating the model's adaptive learning capability. However, removing all expert models (w/o all) leads to increase in error rate, showing that expert models serve as effective prior knowledge for initialization.
\item \textbf{Scenario-specific sensitivity in expert removal.} Eliminating the model collected during walking (w/o w-h) results in incremental error rate escalation in the high RMSA scenario (G2) compared to removing the seated-condition model (w/o s-h). This indicates cross-scenario correlations and shared knowledge representation across different environmental contexts.
\end{itemize}

\subsection{Analysis of Adaptive Weight Learning}
\begin{figure}[t]
  \centering
  \includegraphics[width=\linewidth]{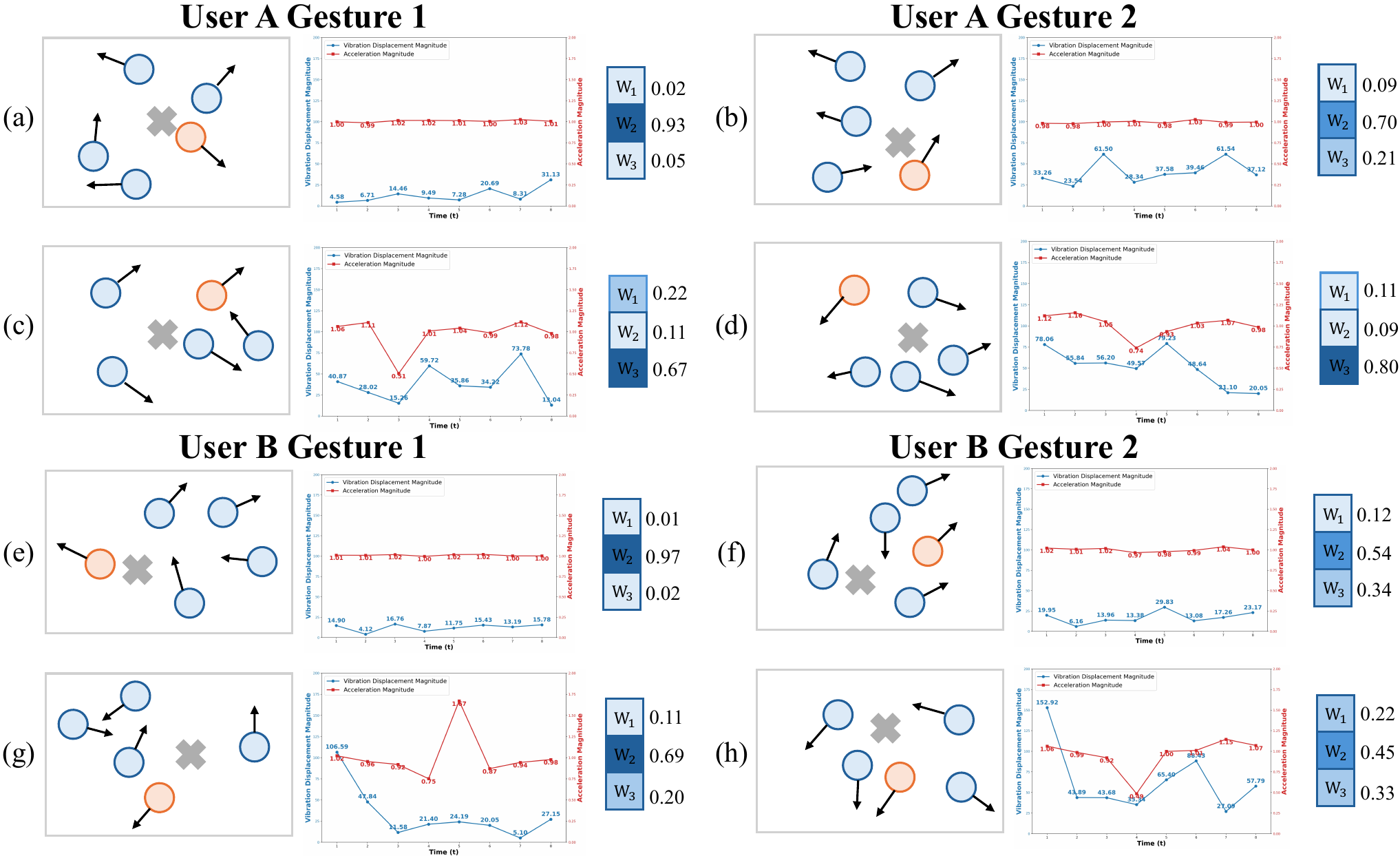}
  \caption{Cases showing how weights are adaptively learned. For each sub-figure, we show a zoom-in scene of moving targets while user touching the screen (left), the trend of accuracy and vibration changing 3 seconds before touching (middle), the fusion weights for experts (right).}
  \label{fig:case}
\end{figure}

To investigate the effect of weight learning on context awareness, we analyzed test results from MAGNeT, trained with 10-shot learning on MTS-2D, focusing on two users in two distinct poses on the test set. As shown in the figure \ref{fig:case}, our observations are as follows:
\begin{itemize}[leftmargin=10pt]
    \item \textbf{Weight adaptation demonstrates well perception of user-specific information.} We observed that for different users in Pose 1 (tablet in hand), the touch points (marked with gray 'X') were generally closer to the actual targets (orange) compared to Pose 2. Specifically, cases (a), (e), and (g) show user selections closer to the target than cases (d), (f), and (h). This is reflected in the fusion weights, where the system assigns higher weights to the expert model representing a seated, handheld gesture.
    \item \textbf{MAGNeT effectively encodes environmental information with stability.} In relatively stable conditions, such as cases (a), (b), (e), and (f), the system consistently assigns higher weights to the second expert model. However, in situations with significant movement or shaking, the weights are adjusted accordingly.
    \item \textbf{Model Adjusts Weights for Complex Environments or High User Uncertainty.} For complex environments or user gestures with high uncertainty, the model dynamically adjusts weights to fuse different expert models. As shown in cases (c) and (h), where acceleration and vibration data exhibit significant fluctuations, the system assigns higher weights to the third expert model, which represents a walking, handheld gesture.
\end{itemize}

\section{Conclusion}
This paper introduces MAGNeT, a novel approach designed to address the challenge of understanding user intent when selecting moving targets. MAGNeT leverages multi-modal information for adaptive learning, significantly reducing error rates by integrating expert models with only a small number of samples. Its effectiveness has been validated through experiments on both 2D and 3D datasets.

However, this study has several limitations that open avenues for future research:
\begin{itemize}[leftmargin=10pt]
    \item The current approach still relies on pre-built user profiles for perceptual understanding. Future work could explore incorporating real-time sensors, such as cameras, to dynamically generate user profiles and extract user features during cold-start scenarios. This would enhance MAGNeT's ability to adapt to new users without prior data.
    \item The current dataset is relatively small, and due to experimental constraints, our vehicular experiments were limited to a single loop-shaped road. This restricts the variety of environmental factors considered. Future research will focus on expanding the dataset to include new road types and diverse environments, such as maritime scenarios (e.g., ship cabins), to further validate MAGNeT's robustness and generalizability.
    \item The current experimental setup did not account for varying numbers of targets. While the principles of Bayesian pointing and MAGNeT's proposed method are theoretically transferable to scenarios with different target quantities, the generalization capability remains unexplored. Future work will investigate MAGNeT's performance and adaptability when the number of moving targets changes.
    
\end{itemize}

\begin{acks}
This work was supported by the National Natural Science Foundation of China (Grant Nos. 62332017, 62172397, 62277035), the Youth Innovation and Technology Support Program of Shandong Provincial Higher Education Institutions under Grant [number 2022KJN028].
\end{acks}

\bibliographystyle{ACM-Reference-Format}
\balance
\bibliography{sample-sigconf}


\begin{thebibliography}{29}

%%% the Web address.

\ifx \showCODEN    \undefined \def \showCODEN     #1{\unskip}     \fi
\ifx \showISBNx    \undefined \def \showISBNx     #1{\unskip}     \fi
\ifx \showISBNxiii \undefined \def \showISBNxiii  #1{\unskip}     \fi
\ifx \showISSN     \undefined \def \showISSN      #1{\unskip}     \fi
\ifx \showLCCN     \undefined \def \showLCCN      #1{\unskip}     \fi
\ifx \shownote     \undefined \def \shownote      #1{#1}          \fi
\ifx \showarticletitle \undefined \def \showarticletitle #1{#1}   \fi
\ifx \showURL      \undefined \def \showURL       {\relax}        \fi
% invisible to TeX
\providecommand\bibfield[2]{#2}
\providecommand\bibinfo[2]{#2}
\providecommand\natexlab[1]{#1}
\providecommand\showeprint[2][]{arXiv:#2}

\bibitem[An and SI(1997)]%
        {an1997mechanical}
\bibfield{author}{\bibinfo{person}{NAIS An} {and} \bibinfo{person}{S SI}.} \bibinfo{year}{1997}\natexlab{}.
\newblock \showarticletitle{Mechanical Vibration and Shock-Evaluation of Human Exposure to Whole-Body Vibration-Part 1: General Requirements}.
\newblock  (\bibinfo{year}{1997}).
\newblock


\bibitem[Cao et~al\mbox{.}(2023)]%
        {cao2023multi}
\bibfield{author}{\bibinfo{person}{Bing Cao}, \bibinfo{person}{Yiming Sun}, \bibinfo{person}{Pengfei Zhu}, {and} \bibinfo{person}{Qinghua Hu}.} \bibinfo{year}{2023}\natexlab{}.
\newblock \showarticletitle{Multi-modal gated mixture of local-to-global experts for dynamic image fusion}. In \bibinfo{booktitle}{\emph{Proceedings of the IEEE/CVF international conference on computer vision}}. \bibinfo{pages}{23555--23564}.
\newblock


\bibitem[Dodd et~al\mbox{.}(2014)]%
        {Dodd2014Touch}
\bibfield{author}{\bibinfo{person}{Sonia Dodd}, \bibinfo{person}{Jeff Lancaster}, \bibinfo{person}{Andrew Miranda}, \bibinfo{person}{Steve Grothe}, \bibinfo{person}{Bob DeMers}, {and} \bibinfo{person}{Bill Rogers}.} \bibinfo{year}{2014}\natexlab{}.
\newblock \showarticletitle{Touch Screens on the Flight Deck: The Impact of Touch Target Size, Spacing, Touch Technology and Turbulence on Pilot Performance}.
\newblock \bibinfo{journal}{\emph{Proceedings of the Human Factors and Ergonomics Society Annual Meeting}} \bibinfo{volume}{58}, \bibinfo{number}{1} (\bibinfo{year}{2014}), \bibinfo{pages}{6--10}.
\newblock
\showeprint{https://doi.org/10.1177/1541931214581002}
\href{https://doi.org/10.1177/1541931214581002}{doi:\nolinkurl{10.1177/1541931214581002}}


\bibitem[Fang and Gerkmann(2023)]%
        {fang2023uncertainty}
\bibfield{author}{\bibinfo{person}{Huajian Fang} {and} \bibinfo{person}{Timo Gerkmann}.} \bibinfo{year}{2023}\natexlab{}.
\newblock \showarticletitle{Uncertainty estimation in deep speech enhancement using complex Gaussian mixture models}. In \bibinfo{booktitle}{\emph{ICASSP 2023-2023 IEEE International Conference on Acoustics, Speech and Signal Processing (ICASSP)}}. IEEE, \bibinfo{pages}{1--5}.
\newblock


\bibitem[Gan et~al\mbox{.}(2025)]%
        {gan2025mixture}
\bibfield{author}{\bibinfo{person}{Wensheng Gan}, \bibinfo{person}{Zhenyao Ning}, \bibinfo{person}{Zhenlian Qi}, {and} \bibinfo{person}{Philip~S Yu}.} \bibinfo{year}{2025}\natexlab{}.
\newblock \showarticletitle{Mixture of Experts (MoE): A Big Data Perspective}.
\newblock \bibinfo{journal}{\emph{arXiv preprint arXiv:2501.16352}} (\bibinfo{year}{2025}).
\newblock


\bibitem[Gershman and Blei(2012)]%
        {gershman2012tutorial}
\bibfield{author}{\bibinfo{person}{Samuel~J Gershman} {and} \bibinfo{person}{David~M Blei}.} \bibinfo{year}{2012}\natexlab{}.
\newblock \showarticletitle{A tutorial on Bayesian nonparametric models}.
\newblock \bibinfo{journal}{\emph{Journal of Mathematical Psychology}} \bibinfo{volume}{56}, \bibinfo{number}{1} (\bibinfo{year}{2012}), \bibinfo{pages}{1--12}.
\newblock


\bibitem[Hasan et~al\mbox{.}(2011)]%
        {Hasan2011Comet}
\bibfield{author}{\bibinfo{person}{Khalad Hasan}, \bibinfo{person}{Tovi Grossman}, {and} \bibinfo{person}{Pourang Irani}.} \bibinfo{year}{2011}\natexlab{}.
\newblock \showarticletitle{Comet and target ghost: techniques for selecting moving targets}. In \bibinfo{booktitle}{\emph{Proceedings of the SIGCHI Conference on Human Factors in Computing Systems}} (Vancouver, BC, Canada) \emph{(\bibinfo{series}{CHI '11})}. \bibinfo{publisher}{Association for Computing Machinery}, \bibinfo{address}{New York, NY, USA}, \bibinfo{pages}{839–848}.
\newblock
\showISBNx{9781450302289}
\href{https://doi.org/10.1145/1978942.1979065}{doi:\nolinkurl{10.1145/1978942.1979065}}


\bibitem[Huang et~al\mbox{.}(2022)]%
        {Huang2022motion-in-depth}
\bibfield{author}{\bibinfo{person}{Jin Huang}, \bibinfo{person}{John~J. Dudley}, \bibinfo{person}{Stephen Uzor}, \bibinfo{person}{Dong Wu}, \bibinfo{person}{Per~Ola Kristensson}, {and} \bibinfo{person}{Feng Tian}.} \bibinfo{year}{2022}\natexlab{}.
\newblock \showarticletitle{Understanding user performance of acquiring targets with motion-in-depth in virtual reality}.
\newblock \bibinfo{journal}{\emph{International Journal of Human-Computer Studies}}  \bibinfo{volume}{163} (\bibinfo{year}{2022}), \bibinfo{pages}{102817}.
\newblock
\showISSN{1071-5819}
\href{https://doi.org/10.1016/j.ijhcs.2022.102817}{doi:\nolinkurl{10.1016/j.ijhcs.2022.102817}}


\bibitem[Huang and Lee(2019)]%
        {huang2019modeling}
\bibfield{author}{\bibinfo{person}{Jin Huang} {and} \bibinfo{person}{Byungjoo Lee}.} \bibinfo{year}{2019}\natexlab{}.
\newblock \showarticletitle{Modeling error rates in spatiotemporal moving target selection}. In \bibinfo{booktitle}{\emph{Extended Abstracts of the 2019 CHI Conference on Human Factors in Computing Systems}}. \bibinfo{pages}{1--6}.
\newblock


\bibitem[Huang et~al\mbox{.}(2020)]%
        {Huang2020Crossing}
\bibfield{author}{\bibinfo{person}{Jin Huang}, \bibinfo{person}{Feng Tian}, \bibinfo{person}{Xiangmin Fan}, \bibinfo{person}{Huawei Tu}, \bibinfo{person}{Hao Zhang}, \bibinfo{person}{Xiaolan Peng}, {and} \bibinfo{person}{Hongan Wang}.} \bibinfo{year}{2020}\natexlab{}.
\newblock \showarticletitle{Modeling the Endpoint Uncertainty in Crossing-Based Moving Target Selection}. In \bibinfo{booktitle}{\emph{Proceedings of the 2020 CHI Conference on Human Factors in Computing Systems}} (Honolulu, HI, USA) \emph{(\bibinfo{series}{CHI '20})}. \bibinfo{publisher}{Association for Computing Machinery}, \bibinfo{address}{New York, NY, USA}, \bibinfo{pages}{1–12}.
\newblock
\showISBNx{9781450367080}
\href{https://doi.org/10.1145/3313831.3376336}{doi:\nolinkurl{10.1145/3313831.3376336}}


\bibitem[Huang et~al\mbox{.}(2018)]%
        {Huang20181D}
\bibfield{author}{\bibinfo{person}{Jin Huang}, \bibinfo{person}{Feng Tian}, \bibinfo{person}{Xiangmin Fan}, \bibinfo{person}{Xiaolong~(Luke) Zhang}, {and} \bibinfo{person}{Shumin Zhai}.} \bibinfo{year}{2018}\natexlab{}.
\newblock \showarticletitle{Understanding the Uncertainty in 1D Unidirectional Moving Target Selection}. In \bibinfo{booktitle}{\emph{Proceedings of the 2018 CHI Conference on Human Factors in Computing Systems}} (Montreal QC, Canada) \emph{(\bibinfo{series}{CHI '18})}. \bibinfo{publisher}{Association for Computing Machinery}, \bibinfo{address}{New York, NY, USA}, \bibinfo{pages}{1–12}.
\newblock
\showISBNx{9781450356206}
\href{https://doi.org/10.1145/3173574.3173811}{doi:\nolinkurl{10.1145/3173574.3173811}}


\bibitem[Huang et~al\mbox{.}(2019)]%
        {Huang20192D}
\bibfield{author}{\bibinfo{person}{Jin Huang}, \bibinfo{person}{Feng Tian}, \bibinfo{person}{Nianlong Li}, {and} \bibinfo{person}{Xiangmin Fan}.} \bibinfo{year}{2019}\natexlab{}.
\newblock \showarticletitle{Modeling the Uncertainty in 2D Moving Target Selection}. In \bibinfo{booktitle}{\emph{Proceedings of the 32nd Annual ACM Symposium on User Interface Software and Technology}} (New Orleans, LA, USA) \emph{(\bibinfo{series}{UIST '19})}. \bibinfo{publisher}{Association for Computing Machinery}, \bibinfo{address}{New York, NY, USA}, \bibinfo{pages}{1031–1043}.
\newblock
\showISBNx{9781450368162}
\href{https://doi.org/10.1145/3332165.3347880}{doi:\nolinkurl{10.1145/3332165.3347880}}


\bibitem[Ilich(2009)]%
        {ilich2009moving}
\bibfield{author}{\bibinfo{person}{Michael~Victor Ilich}.} \bibinfo{year}{2009}\natexlab{}.
\newblock \emph{\bibinfo{title}{Moving target selection in interactive video}}.
\newblock \bibinfo{thesistype}{Ph.\,D. Dissertation}. \bibinfo{school}{University of British Columbia}.
\newblock


\bibitem[Kim and Martin(2013)]%
        {kim2013biodynamic}
\bibfield{author}{\bibinfo{person}{Heon-Jeong Kim} {and} \bibinfo{person}{Bernard~J Martin}.} \bibinfo{year}{2013}\natexlab{}.
\newblock \showarticletitle{Biodynamic characteristics of upper limb reaching movements of the seated human under whole-body vibration}.
\newblock \bibinfo{journal}{\emph{Journal of applied biomechanics}} \bibinfo{volume}{29}, \bibinfo{number}{1} (\bibinfo{year}{2013}), \bibinfo{pages}{12--22}.
\newblock


\bibitem[Li et~al\mbox{.}(2023)]%
        {li2023adaptive}
\bibfield{author}{\bibinfo{person}{Jiamin Li}, \bibinfo{person}{Qiang Su}, \bibinfo{person}{Yitao Yang}, \bibinfo{person}{Yimin Jiang}, \bibinfo{person}{Cong Wang}, {and} \bibinfo{person}{Hong Xu}.} \bibinfo{year}{2023}\natexlab{}.
\newblock \showarticletitle{Adaptive gating in mixture-of-experts based language models}.
\newblock \bibinfo{journal}{\emph{arXiv preprint arXiv:2310.07188}} (\bibinfo{year}{2023}).
\newblock


\bibitem[Li et~al\mbox{.}(2022)]%
        {Li2022Select}
\bibfield{author}{\bibinfo{person}{Zhi Li}, \bibinfo{person}{Maozheng Zhao}, \bibinfo{person}{Dibyendu Das}, \bibinfo{person}{HANG ZHAO}, \bibinfo{person}{Yan Ma}, \bibinfo{person}{Wanyu Liu}, \bibinfo{person}{Michel Beaudouin-Lafon}, \bibinfo{person}{Fusheng Wang}, \bibinfo{person}{IV Ramakrishnan}, {and} \bibinfo{person}{Xiaojun Bi}.} \bibinfo{year}{2022}\natexlab{}.
\newblock \showarticletitle{Select or Suggest? Reinforcement Learning-based Method for High-Accuracy Target Selection on Touchscreens}. In \bibinfo{booktitle}{\emph{Proceedings of the 2022 CHI Conference on Human Factors in Computing Systems}} (New Orleans, LA, USA) \emph{(\bibinfo{series}{CHI '22})}. \bibinfo{publisher}{Association for Computing Machinery}, \bibinfo{address}{New York, NY, USA}, Article \bibinfo{articleno}{494}, \bibinfo{numpages}{15}~pages.
\newblock
\showISBNx{9781450391573}
\href{https://doi.org/10.1145/3491102.3517472}{doi:\nolinkurl{10.1145/3491102.3517472}}


\bibitem[Lu et~al\mbox{.}(2020)]%
        {lu2020investigating}
\bibfield{author}{\bibinfo{person}{Yiqin Lu}, \bibinfo{person}{Chun Yu}, {and} \bibinfo{person}{Yuanchun Shi}.} \bibinfo{year}{2020}\natexlab{}.
\newblock \showarticletitle{Investigating bubble mechanism for ray-casting to improve 3D target acquisition in virtual reality}. In \bibinfo{booktitle}{\emph{2020 IEEE Conference on Virtual Reality and 3D User Interfaces (VR)}}. IEEE, \bibinfo{pages}{35--43}.
\newblock


\bibitem[Mankoff et~al\mbox{.}(2000)]%
        {Mankoff2000Providing}
\bibfield{author}{\bibinfo{person}{Jennifer Mankoff}, \bibinfo{person}{Scott~E. Hudson}, {and} \bibinfo{person}{Gregory~D. Abowd}.} \bibinfo{year}{2000}\natexlab{}.
\newblock \showarticletitle{Providing integrated toolkit-level support for ambiguity in recognition-based interfaces}. In \bibinfo{booktitle}{\emph{Proceedings of the SIGCHI Conference on Human Factors in Computing Systems}} (The Hague, The Netherlands) \emph{(\bibinfo{series}{CHI '00})}. \bibinfo{publisher}{Association for Computing Machinery}, \bibinfo{address}{New York, NY, USA}, \bibinfo{pages}{368–375}.
\newblock
\showISBNx{1581132166}
\href{https://doi.org/10.1145/332040.332459}{doi:\nolinkurl{10.1145/332040.332459}}


\bibitem[Mansfield and Maeda(2005)]%
        {mansfield2005effect}
\bibfield{author}{\bibinfo{person}{Neil~J Mansfield} {and} \bibinfo{person}{Setsuo Maeda}.} \bibinfo{year}{2005}\natexlab{}.
\newblock \showarticletitle{Effect of backrest and torso twist on the apparent mass of the seated body exposed to vertical vibration}.
\newblock \bibinfo{journal}{\emph{Industrial Health}} \bibinfo{volume}{43}, \bibinfo{number}{3} (\bibinfo{year}{2005}), \bibinfo{pages}{413--420}.
\newblock


\bibitem[McLachlan(2000)]%
        {mclachlan2000finite}
\bibfield{author}{\bibinfo{person}{Geoffrey McLachlan}.} \bibinfo{year}{2000}\natexlab{}.
\newblock \showarticletitle{Finite mixture models}.
\newblock \bibinfo{journal}{\emph{A wiley-interscience publication}} (\bibinfo{year}{2000}).
\newblock


\bibitem[Schneider and Graham(2023)]%
        {schneider2023supporting}
\bibfield{author}{\bibinfo{person}{Adrian L~Jessup Schneider} {and} \bibinfo{person}{TC~Nicholas Graham}.} \bibinfo{year}{2023}\natexlab{}.
\newblock \showarticletitle{Supporting aim assistance algorithms through a rapidly trainable, personalized model of players’ spatial and temporal aiming ability}. In \bibinfo{booktitle}{\emph{Proceedings of the 2023 CHI Conference on Human Factors in Computing Systems}}. \bibinfo{pages}{1--17}.
\newblock


\bibitem[Shadmehr et~al\mbox{.}(2010)]%
        {shadmehr2010error}
\bibfield{author}{\bibinfo{person}{Reza Shadmehr}, \bibinfo{person}{Maurice~A Smith}, {and} \bibinfo{person}{John~W Krakauer}.} \bibinfo{year}{2010}\natexlab{}.
\newblock \showarticletitle{Error correction, sensory prediction, and adaptation in motor control}.
\newblock \bibinfo{journal}{\emph{Annual review of neuroscience}} \bibinfo{volume}{33}, \bibinfo{number}{1} (\bibinfo{year}{2010}), \bibinfo{pages}{89--108}.
\newblock


\bibitem[Vanacken et~al\mbox{.}(2007)]%
        {Vanacken2007BubbleCursor}
\bibfield{author}{\bibinfo{person}{Lode Vanacken}, \bibinfo{person}{Tovi Grossman}, {and} \bibinfo{person}{Karin Coninx}.} \bibinfo{year}{2007}\natexlab{}.
\newblock \showarticletitle{Exploring the Effects of Environment Density and Target Visibility on Object Selection in 3D Virtual Environments}. In \bibinfo{booktitle}{\emph{2007 IEEE Symposium on 3D User Interfaces}}.
\newblock
\href{https://doi.org/10.1109/3DUI.2007.340783}{doi:\nolinkurl{10.1109/3DUI.2007.340783}}


\bibitem[Yi and Chen(2024)]%
        {yi2024variational}
\bibfield{author}{\bibinfo{person}{Jing Yi} {and} \bibinfo{person}{Zhenzhong Chen}.} \bibinfo{year}{2024}\natexlab{}.
\newblock \showarticletitle{Variational Mixture of Stochastic Experts Auto-encoder for Multi-modal Recommendation}.
\newblock \bibinfo{journal}{\emph{IEEE Transactions on Multimedia}} (\bibinfo{year}{2024}).
\newblock


\bibitem[Zhang et~al\mbox{.}(2023)]%
        {zhang2023shape}
\bibfield{author}{\bibinfo{person}{Hao Zhang}, \bibinfo{person}{Jin Huang}, \bibinfo{person}{Huawei Tu}, {and} \bibinfo{person}{Feng Tian}.} \bibinfo{year}{2023}\natexlab{}.
\newblock \showarticletitle{Shape-Adaptive Ternary-Gaussian Model: Modeling Pointing Uncertainty for Moving Targets of Arbitrary Shapes}. In \bibinfo{booktitle}{\emph{Proceedings of the 2023 CHI Conference on Human Factors in Computing Systems}}. \bibinfo{pages}{1--18}.
\newblock


\bibitem[Zhang et~al\mbox{.}(2020)]%
        {Zhang2020shape}
\bibfield{author}{\bibinfo{person}{Ziyue Zhang}, \bibinfo{person}{Jin Huang}, {and} \bibinfo{person}{Feng Tian}.} \bibinfo{year}{2020}\natexlab{}.
\newblock \showarticletitle{Modeling the Uncertainty in Pointing of Moving Targets with Arbitrary Shapes}. In \bibinfo{booktitle}{\emph{Extended Abstracts of the 2020 CHI Conference on Human Factors in Computing Systems}} (Honolulu, HI, USA) \emph{(\bibinfo{series}{CHI EA '20})}. \bibinfo{publisher}{Association for Computing Machinery}, \bibinfo{address}{New York, NY, USA}, \bibinfo{pages}{1–7}.
\newblock
\showISBNx{9781450368193}
\href{https://doi.org/10.1145/3334480.3382875}{doi:\nolinkurl{10.1145/3334480.3382875}}


\bibitem[zheng et~al\mbox{.}(2021)]%
        {zheng2021AScenario}
\bibfield{author}{\bibinfo{person}{yawen zheng}, \bibinfo{person}{Jin Huang}, \bibinfo{person}{Juan Liu}, \bibinfo{person}{Chenglei Yang}, {and} \bibinfo{person}{Feng Tian}.} \bibinfo{year}{2021}\natexlab{}.
\newblock \showarticletitle{A Scenario Adaptive Model for Predicting Error Rates in Moving Target Selection on Smartphones}. In \bibinfo{booktitle}{\emph{Proceedings of the 23rd International Conference on Mobile Human-Computer Interaction}} (Toulouse \&amp; Virtual, France) \emph{(\bibinfo{series}{MobileHCI '21})}. \bibinfo{publisher}{Association for Computing Machinery}, \bibinfo{address}{New York, NY, USA}, Article \bibinfo{articleno}{10}, \bibinfo{numpages}{15}~pages.
\newblock
\showISBNx{9781450383288}
\href{https://doi.org/10.1145/3447526.3472049}{doi:\nolinkurl{10.1145/3447526.3472049}}


\bibitem[Zheng et~al\mbox{.}(2025)]%
        {Zheng3D}
\bibfield{author}{\bibinfo{person}{Yawen Zheng}, \bibinfo{person}{Jin Huang}, \bibinfo{person}{Hao Zhang}, \bibinfo{person}{Yulong Bian}, \bibinfo{person}{Juan Liu}, \bibinfo{person}{Chenglei Yang}, \bibinfo{person}{Feng Tian}, {and} \bibinfo{person}{Xiangxu Meng}.} \bibinfo{year}{2025}\natexlab{}.
\newblock \showarticletitle{3D Ternary-Gaussian model: Modeling pointing uncertainty of 3D moving target selection in virtual reality}.
\newblock \bibinfo{journal}{\emph{International Journal of Human-Computer Studies}}  \bibinfo{volume}{198} (\bibinfo{year}{2025}), \bibinfo{pages}{103454}.
\newblock
\showISSN{1071-5819}
\href{https://doi.org/10.1016/j.ijhcs.2025.103454}{doi:\nolinkurl{10.1016/j.ijhcs.2025.103454}}


\bibitem[Zhu et~al\mbox{.}(2020)]%
        {ZhuBayesCommand}
\bibfield{author}{\bibinfo{person}{Suwen Zhu}, \bibinfo{person}{Yoonsang Kim}, \bibinfo{person}{Jingjie Zheng}, \bibinfo{person}{Jennifer~Yi Luo}, \bibinfo{person}{Ryan Qin}, \bibinfo{person}{Liuping Wang}, \bibinfo{person}{Xiangmin Fan}, \bibinfo{person}{Feng Tian}, {and} \bibinfo{person}{Xiaojun Bi}.} \bibinfo{year}{2020}\natexlab{}.
\newblock \showarticletitle{Using Bayes' Theorem for Command Input: Principle, Models, and Applications}. In \bibinfo{booktitle}{\emph{Proceedings of the 2020 CHI Conference on Human Factors in Computing Systems}} (Honolulu, HI, USA) \emph{(\bibinfo{series}{CHI '20})}. \bibinfo{publisher}{Association for Computing Machinery}, \bibinfo{address}{New York, NY, USA}, \bibinfo{pages}{1–15}.
\newblock
\showISBNx{9781450367080}
\href{https://doi.org/10.1145/3313831.3376771}{doi:\nolinkurl{10.1145/3313831.3376771}}


\end{thebibliography}

\end{document}